\documentclass[%
preprint,
 amsmath,amssymb,
 aps, nofootinbib
]{revtex4-2}

\usepackage{graphicx}
\usepackage{dcolumn}
\usepackage{bm}
\usepackage{physics}
\usepackage{xcolor}
\usepackage{appendix}

\usepackage{caption}
\usepackage{subcaption}

\usepackage{bbold}
\begin{document}


\title{Gravitational Harmonium: Gravitationally Induced Entanglement in a Harmonic Trap}


\author{Jackson Yant}%
\email{Jackson.R.Yant.Gr@Dartmouth.edu}
\affiliation{Department of Physics and Astronomy, Dartmouth College, Hanover, New Hampshire 03755, USA}


\author{Miles Blencowe}
\email{Miles.P.Blencowe@Dartmouth.edu}
\affiliation{Department of Physics and Astronomy, Dartmouth College, Hanover, New Hampshire 03755, USA}


\date{\today}


\begin{abstract}
    Recent work has shown that it may be possible to detect gravitationally induced entanglement in tabletop experiments in the not-too-distant future. However, there are at present no thoroughly developed models for this type of experiment where the entangled particles are treated more fundamentally as excitations of a relativistic quantum field, and with the measurements modeled using expectation values of field observables. Here we propose a thought experiment where two particles (i.e., massive scalar field quanta) are initially prepared in a superposition of coherent states within a common three-dimensional (3D) harmonic trap. The particles then develop entanglement through their mutual gravitational interaction, which can be probed through particle position detection probabilities. The present work gives a non-relativistic quantum mechanical analysis of the gravitationally induced entanglement of this system, which we term the `gravitational harmonium' due to its similarity to the harmonium model of approximate electron interactions in a helium atom; the entanglement is operationally determined through the matter wave interference visibility. The present work serves as the basis for a subsequent investigation, which models this system using quantum field theory, providing further insights into the quantum nature of gravitationally induced entanglement through  relativistic corrections, together with an operational procedure to quantify the entanglement.
\end{abstract}
\maketitle
\newpage
\section{\label{introduction} Introduction}

Quantum mechanics and general relativity are our two most fundamental, experimentally tested, and predictive theories.  However, these theories are mutually incompatible and no comparably successful  theory unifying the two has been formulated, while little experimental progress has been made. Recently, Bose {\it et al}. \cite{bose2017} and Marletto and Vedral \cite{Marletto:2017kzi} (BMV) proposed  in-principle-realizable experiments in which two particles could become entangled by their mutual gravitational interaction \cite{bose2017,Marletto:2017kzi}. Detection of gravitationally induced entanglement would then provide novel empirical evidence that might guide the development of new theories aiming at resolving the tensions between classical gravity and quantum mechanics.

The BMV proposals have inspired a great deal of interest in the quantum information science community, with ongoing discussions about the theoretical implications of detecting gravitationally induced entanglement. Some contend that a positive detection would provide definitive evidence for the quantum nature of the gravitational field \cite{bose2017,Marletto:2017kzi,Marshman2020,Marletto2019,marletto2018,Galley2020,Marletto2019, Marletto2020,CHRISTODOULOU2019, Danielson2022}. Others have pointed out possible ways in which a classical model of gravity could induce entanglement in such an experiment \cite{Hall2018,Reginatto2018,rydving_gedanken_2021,anastopoulos2018comment, Anastopoulos_2021, ma_limits_2022,fragkos_inference_2022}. Some have raised points about how several of the principles of quantum information theory used to assert that gravitationally induced entanglement necessitates quantum gravity, are not clearly satisfied in a quantum field theoretic setting \cite{Altamirano_2018,anastopoulos_quantum_2021,anastopoulos_quantum_2022,Anastopoulos2022}. In response, there have also been several investigations into what implications can be drawn from gravitationally induced entanglement without necessarily relying on such assumptions \cite{christodoulou2022,christodoulou_gravity_2022,martin-martinez_what_2022,delisle_quantum_nodate}.

Several variations of a gravitationally induced entanglement experiment have been proposed \cite{Nguyen:2019huk,Tilly2021,Krisnanda:2019glc,vandeKamp2020,matsumura2020,chevalier2020,Carney2021,Carlesso_2019,Feng:2022hfv,zhou_catapulting_2022,marshman_constructing_2022,zhou_gravito-diamagnetic_2022,schut_improving_2022,guff_optimal_2022}. To date, none of these proposed experiments have been modelled using a  fully quantum field theoretic description of both gravity, the  particles involved and the measures of entanglement generated between the latter. Such an {\it{ab initio}} approach would allow one to explore the quantum nature of the gravitational field, including relativistic effects \cite{Carney2022}. In this work, we seek to develop a  gravitationally induced entanglement thought experiment that adapts the BMV proposal to enable such a description. 

\begin{figure}
    \centering
    \includegraphics[scale=0.26]{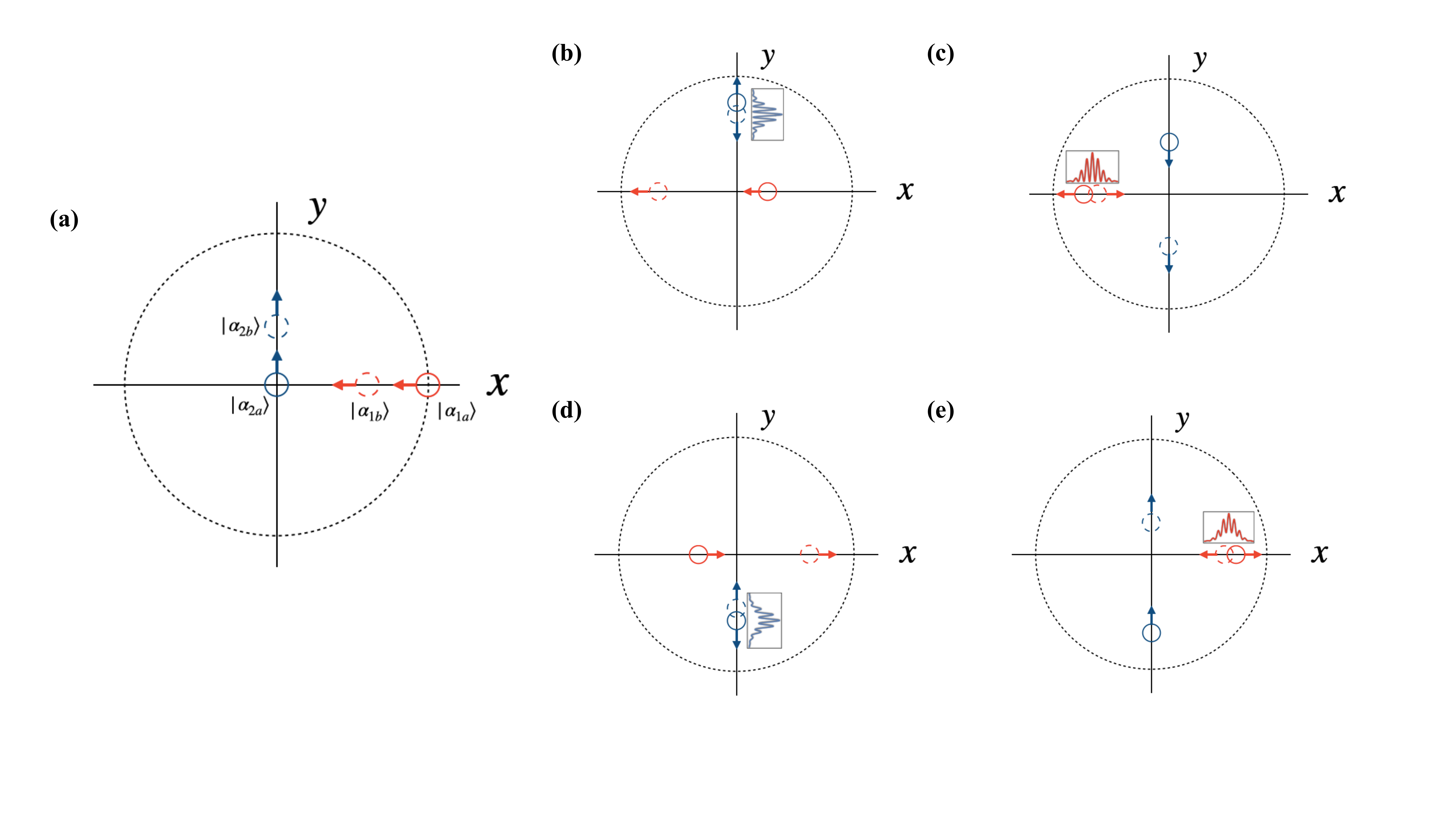}
    \caption{Scheme of the thought experiment for generating and probing gravitationally-induced entanglement between two massive particles. Snapshots (a)-(e) show the coherent state position maxima of the two particles, each initially in a superposition state, at subsequent times. When the particle wavefunction components in a given superposition state overlap, interference fringes develop in the spatial dependence of the particle detection probability along the oscillation axis of the particle. The fringe visibility decreases over time, corresponding to increasing entanglement between the two particles.}
    \label{fig:setup}
\end{figure}
Our thought experiment considers two particles that are located within a 3D harmonic trap. The initial condition is a product state wherein each particle is prepared in a displaced coherent state superposition along  mutually orthogonal axes. As the coherent state wavefunction components of one of the particles evolve, they will overlap  periodically, producing interference fringes in the spatial probability distribution for detecting the particle. The gravitationally-induced entanglement that develops between the two particles will cause the visibility of these fringes to decrease. Figure \ref{fig:setup} illustrates this scheme. We term this model  the `gravitational harmonium' after its similarity to a model for approximating the Helium electrons' Coulombic trapping potentials with harmonic potentials \cite{cioslowski2000}.\footnote{While we view such a scheme as a thought experiment that allows for a relativistic quantum field theoretic analysis, it does resemble recent optical trapping schemes for massive particles that share a similar goal: to probe gravitational entanglement in quantum systems \cite{delic2020}.}

The present work analyzes the gravitational harmonium using non-relativistic quantum mechanics, serving as the basis for a future investigation that will apply a quantum field theory analysis to the same system \cite{yant2023}. In particular, the gravitational harmonium is readily suited for a full, quantum field theoretic treatment. Under low energy laboratory conditions, we can model such a thought experiment by a massive real scalar field $\phi$ coupled to gravity through the Einstein Hilbert action, expanded to second order in $\kappa=\sqrt{32\pi G}$ (with units $\hbar=c=1$)  \cite{donoghue1994,arteaga2004}:
\begin{equation}
    S[\phi,h_{\mu\nu}]=S_M[\phi]+S_E[h_{\mu\nu}]+S_I[\phi,h_{\mu\nu}].
    \label{sysenveq}
\end{equation} 
The action for the scalar field, $\phi$, contains a  harmonic potential that is spatially centered, e.g., at rectilinear coordinate location ${\bf{r}}=0$ and given by
\begin{equation}
    S_M[\phi]=-\frac{1}{2}\int d^4x\, \left[\eta^{\mu\nu}\partial_{\mu} \phi\partial_{\nu}\phi+m^2\left(1+\omega^2r^2\right)\phi^2\right],
    \label{matteracteq}
\end{equation}
while the gravity and interaction actions are respectively
\begin{equation}
\begin{split}
    S_E[h_{\mu\nu}]=&\int d^4 x\left(-\frac{1}{2} \partial^{\rho} h^{\mu\nu} \partial_{\rho} h_{\mu\nu}+\partial_{\nu} h^{\mu\nu}\partial^{\rho}h_{\mu\rho}-\partial_{\mu}h \partial_{\nu}h^{\mu\nu}+\frac{1}{2}\partial^{\mu}h\partial_{\mu}h\right),\\
    S_{{I}}=&\int d^4 x\left(\frac{\kappa}{2} T^{\mu\nu}\left(\phi\right)h_{\mu\nu}+ \frac{\kappa^2}{4}U^{\mu\nu\rho\sigma}\left(\phi\right)h_{\mu\nu}h_{\rho\sigma}\right),
\end{split}
\label{gravenveq}
\end{equation}
where $T_{\mu\nu}\left(\phi\right)$ is the scalar field energy-momentum tensor,  and $U_{\mu\nu\rho\sigma}\left(\phi\right)$ is a quadratic tensor in $\phi$ \cite{arteaga2004}. Decoherence aspects of certain states within such a framework have been considered in Ref. \cite{oniga2017}, and in a  0D model \cite{xu_zero-dimensional_2022}. We may then describe the fringe visibility by using the expectation value of the particle detection probability given in the Schr\"{o}dinger picture,
 \begin{equation}
    {\mathrm{Tr}}\left[\rho\left(t\right) \left(V^{-1}\int_V d{\bf{x}} \phi\left({\bf{x}},0\right)\right)^2  \right]=V^{-2}\int_V d{\bf{x}}d{}{\bf x}'{\mathrm{Tr}}\left[\rho\left(t\right)\phi\left({\bf{x}},0\right)\phi\left({\bf{x}}',0\right)\right],
    \label{detectoraveq}
\end{equation}
where $\rho\left(t\right)$ is the density operator for the evolving two-particle state and $V$ is some small coordinate averaging volume corresponding to the spatial region occupied by the particle detector.

In Section \ref{sec:harmonium} we solve the corresponding non-relativistic Schr\"{o}dinger equation to derive the approximate time evolution of the two-particle, initial superposition pair state of this system as gravitational entanglement between the two particles develops, with an accompanying fringe visibility reduction in the spatial dependence of the single particle detection probability; further details are given in three appendices. As discussed above, this analysis will be used as a reference for future work where the corresponding quantum dynamics will be obtained within the above-described quantum field theoretic  description \cite{yant2023}; we hope to gain new insights into how gravity induces quantum entanglement by comparing the approximate non-relativistic model developed here with the more fundamental one involving relativistic quantum fields. Section \ref{sec:potentials} considers the effect that alternative, modified gravity interaction potentials would have on the non-relativistic quantum mechanical model for entanglement generation and the corresponding reduction in visibility. Section \ref{sec:conclusion} gives some concluding remarks.

\section{\label{sec:harmonium}NRQM Gravitational Harmonium}
Consider two particles with identical mass $m$ that are confined to a harmonic trap with frequency $\omega$ and interacting through gravity. The non-relativistic quantum dynamics of this system is described by the  Schr\"odinger equation with Hamiltonian given by
\begin{equation}
    H=\frac{p_1^2}{2m}+\frac{p_2^2}{2m}+\frac{1}{2}m\omega^2x_1^2+\frac{1}{2}m\omega^2x_2^2-\frac{Gm^2}{|\bm{x}_1-\bm{x}_2|},
    \label{Hameq}
\end{equation}
where $\bm{x}_1$ and $\bm{x}_2$ are the position coordinates of particles 1 and 2, respectively.  This Hamiltonian is qualitatively similar to that of the Helium atom  [with the substitution $Gm^2\rightarrow-e^2$ in Eq. (\ref{Hameq})], where the electrons' Coulomb confining potentials are approximately replaced by harmonic confining potentials, and is known as `Hooke's atom' or the `Harmonium' \cite{cioslowski2000}. Because of this similarity, we name this model the `gravitational harmonium.'  Hamiltonian (\ref{Hameq}) can be conveniently  expressed in dimensionless form with the position and time coordinate subsitutions $x_i\rightarrow\sqrt{\frac{m\omega}{\hbar}}x_i$,  $t\rightarrow\omega t$, and $H\rightarrow H/\hbar\omega$, giving:
\begin{equation}
    H=\frac{1}{2}\left(p_1^2+p_2^2+x_1^2+x_2^2\right)-\frac{\sqrt{2}g_0}{|\bm{x}_1-\bm{x}_2|},
    \label{Ham2eq}
\end{equation}
where $g_0=G\sqrt{\frac{m^5}{2\hbar^3\omega}}$ is a dimensionless parameter characterizing the Newtonian gravitational interaction between the two particles. In these scaled, dimensionless coordinates, the Schr\"odinger equation is given as $i\partial_{t}\psi=H\psi$.

The initial state of our system is taken to be a product state wherein particle 1 is in a superposition of coherent states $|\bm{\alpha}_1\rangle$ displaced along the $x$ axis direction [i.e., $\bm{\alpha}_1=(\alpha_{1 x},0,0)$, with $\alpha_{1 x}\neq 0$], 
while particle 2 is in a superposition of coherent states $|\bm{\alpha}_2\rangle$ displaced along the orthogonal $y$ axis direction [i.e., $\bm{\alpha}_2=(0,\alpha_{2 y},0)$, with $\alpha_{2 y}\neq 0$]   
(Fig. \ref{fig:setup}):
\begin{equation}
\begin{split}
    \ket{\psi}=&N\left(\ket{\bm{\alpha}_{1a}}+\ket{\bm{\alpha}_{1b}}\right)\left(\ket{\bm{\alpha}_{2a}}+\ket{\bm{\alpha}_{2b}}\right) \\
    =&N\left(\ket{\bm{\alpha}_{1a},\,\bm{\alpha}_{2a}}+\ket{\bm{\alpha}_{1a},\,\bm{\alpha}_{2b}}+\ket{\bm{\alpha}_{1b},\,\bm{\alpha}_{2a}}+\ket{\bm{\alpha}_{1b},\,\bm{\alpha}_{2b}}\right),
\end{split}
\label{initeq}
\end{equation}
where $N$ is a normalization factor. Here, the single particle coherent states are each parameterized by a complex vector $\bm{\alpha}=\left(\expval{\bm{x}}+i\expval{\bm{p}}\right)/\sqrt{2}$. For the initial state (\ref{initeq}), both particles subsequently remain undisplaced in the $z$-coordinate axis direction and we therefore suppress further mention of the $z$ coordinate and treat $\bm{x}_1$ and $\bm{x}_2$ as two dimensional vectors from now on; the  coherent state parameter vectors $\bm{\alpha}$ are then given by a pair of $x$ and $y$ component complex numbers for each particle [$\bm{\alpha}_1=(\alpha_{1 x},0)$ and $\bm{\alpha}_2=(0,\alpha_{2 y})$]. 

In order to determine the subsequent Schr\"{o}dinger evolution of the initial state in Eq. (\ref{initeq}) with Hamiltonian (\ref{Ham2eq}), we first consider the evolution of one of the individual coherent state products $\ket{\bm{\alpha}_1,\,\bm{\alpha}_2}$ for particles 1 and 2 that make up the superposition on the second line of Eq. (\ref{initeq}). We can then determine the full state evolution by linearity.
It is convenient to work in terms of symmetric center of mass (SCOM) coordinates given by
\begin{equation}
    \bm{R}=\frac{1}{\sqrt{2}}\left(\bm{x}_1+\bm{x}_2\right) \ \mathrm{and,}\  \bm{r}=\frac{1}{\sqrt{2}}\left(\bm{x}_1-\bm{x}_2\right),
\end{equation}
with $\bm{R}$ referred to as the center of mass coordinate and $\bm{r}$ the relative difference coordinate. In terms of these coordinates,  Hamiltonian (\ref{Ham2eq}) decomposes into the sum of the center of mass and relative difference coordinate Hamiltonians:
\begin{equation}
        H=\frac{1}{2}\left(p_R^2+R^2\right) +\frac{1}{2}\left(p_r^2 +r^2-\frac{g_0}{r}\right).
        \label{scomhameq}
\end{equation}
The ket $\ket{\bm{\alpha}_1,\,\bm{\alpha}_2}$ can equivalently be described by coherent state parameter vectors for the SCOM coordinates as follows:
\begin{equation}
    \bm{\alpha}_R=\frac{1}{\sqrt{2}}\left(\bm{\alpha}_1+\bm{\alpha}_2\right)\ \mathrm{and,}\  \bm{\alpha}_r=\frac{1}{\sqrt{2}}\left(\bm{\alpha}_1-\bm{\alpha}_2\right).
\end{equation}
These parameter vectors are nonzero in both their $x$ and $y$ components. From Eq. (\ref{scomhameq}), the Hamiltonian is separable in the SCOM coordinates, and the center of mass coordinate evolves as a simple, 2D quantum harmonic oscillator.

In order to approximate the evolution of the relative difference coordinate, we Taylor expand the potential terms in our Hamiltonian to second order about the position expectation values. For coherent states, whose wavefunctions are given by Gaussians of the form
\begin{equation}
    \psi\left(\bm{r}\right)=\exp i\left[\left(\bm{r}-\expval{\bm{r}}\right)^T\Omega\left(\bm{r}-\expval{\bm{r}}\right)+\expval{\bm{p}_r}\left(\bm{r}-\expval{\bm{r}}\right)+\gamma\right],
\end{equation}
the Gaussian parameters then evolve as follows \cite{heller_1975}:
\begin{align}
    \expval{\dot{\bm{r}}}&=\frac{\partial H}{\partial \expval{\bm{p}_r}},\quad\expval{\dot{\bm{p}_r}}=-\frac{\partial H}{\partial \expval{\bm{r}}},\label{eqn:Heller1}\\
    \dot{\Omega}&=-2\Omega^2-\frac{1}{2}\frac{\partial^2 V}{\partial r_i \partial r_j}\Big|_{\bm{r}=\expval{\bm{r}}},\label{eqn:Heller2}\\
    \dot{\gamma}&=i\, \mathrm{Tr}\,\Omega+\mathcal{L}.
    \label{eqn:Heller3}
\end{align}
Here, $\mathcal{L}$ is the system's classical Lagrangian evaluated for the solutions $\expval{\bm{r}}$ and $\expval{\bm{p}_r}$, and $V$ includes the harmonic and gravitational potential terms. We consider initial coherent state parameter values  $\bm{\alpha}_1=(\alpha,0)$ and $\bm{\alpha}_2=(0,i\alpha)$ for some real number $\alpha$. This corresponds to the condition that the average relative distance  $\expval{\bm{r}}$ between  particles 1 and 2 remains fixed in the absence of their mutual gravitational attraction  ($g_0=0$). Approximate solutions to  Eqs. (\ref{eqn:Heller1})-(\ref{eqn:Heller3}) are derived in Appendix \ref{app:heller apx}, and their numerical validation  is given in Appendix \ref{app:numerics}.

As an aside, it is interesting to note that the initial product state $\ket{\bm{\alpha}_1,\bm{\alpha}_2}$ evolves into a state where particles 1 and 2 are entangled due to nonzero off-diagonal terms in $\Omega$. The entanglement entropy is approximately
\begin{align}
    S\approx\frac{9 g_0^2 t^2 }{64 \alpha^6}\left[1-\log \left(\frac{9 g_0^2 t^2}{128 \alpha^6}\right)\right],
    \label{eqn:covent}
\end{align}
for $\frac{g_0t}{\alpha^3}\ll1$.
This is in contrast to other gravitational entanglement proposals, where the approximate time-evolved state only develops entanglement due to the different gravitationally-induced relative phases between various  localized position state components resulting from their initial spatial superpositions. Such entanglement results in our method of approximation since it keeps track of the motion of the particles' expectation values subject to the gravitational interaction between them. 
However, the gravitationally-induced relative phase terms will provide the dominant contribution to the entanglement between the two particles when we consider superpositions of spatially localized states as we show in the following.

 In order to isolate the entanglement arising from the gravitationally-induced relative phases, we only keep the time-dependent phase contributions to each initial product state component of the full superposition state that arise  from evaluating the action along the gravitationally perturbed two-particle trajectories: 
\begin{align}
     \ket{\psi}=&N\left(e^{i\delta S_{aa}}\ket{\bm{\alpha}_{1a},\,\bm{\alpha}_{2a}}+e^{i\delta S_{ab}}\ket{\bm{\alpha}_{1a},\,\bm{\alpha}_{2b}}+e^{i\delta S_{ba}}\ket{\bm{\alpha}_{1b},\,\bm{\alpha}_{2a}}+e^{i\delta S_{bb}}\ket{\bm{\alpha}_{1b},\,\bm{\alpha}_{2b}}\right),
     \label{approxstateq}
\end{align}
where $N$ is the normalization. The $\delta S$ terms  are given by the differences between the action along the classical path for particles subject to both the harmonic and gravitational interactions and the action for the harmonic  interaction alone, where the initial conditions are given by the position and momentum expectation values for the coherent product state component the $e^{i \delta S}$ phase terms are each multiplying. 

We choose the following initial coherent state parameter values related to the individual particle (not SCOM) coordinates: $\alpha_{1ax}=\alpha,\alpha_{1bx}=\alpha e^{-i\pi/4},\alpha_{2ay}=\alpha e^{i\pi/2},$ and $\alpha_{2by}=\alpha e^{i\pi/4}$, with $\alpha$ a real number [see Fig. \ref{fig:setup}(a)]. This corresponds to components $\ket{\bm{\alpha}_{1a},\,\bm{\alpha}_{2a}}$ and $\ket{\bm{\alpha}_{1b},\,\bm{\alpha}_{2b}}$ both being in the configuration described for the single coherent product state component above where the particles' relative difference coordinate is constant in our approximation. These two pairs are out of phase with each other by a phase factor of $\pi/4$. This implies that the components $\ket{\bm{\alpha}_{1a},\,\bm{\alpha}_{2b}}$ and $\ket{\bm{\alpha}_{1b},\,\bm{\alpha}_{2a}} $ have coherent state parameter values for the $r_x$ and $r_y$ components that are out of phase by $\pm\pi/4$, which does not correspond exactly to the evolution considered for the single component case above, but may be approximated by the same methods as described in Appendix \ref{app:heller apx}. The entanglement entropy for the state (\ref{approxstateq}) is approximated in Appendix \ref{app:phase entropy} and we obtain
\begin{align}
   S\approx&\left(\frac{c\,g_0 }{\alpha}t\right)^2\left\{1-\log\left[\left(\frac{c\,g_0}{\alpha}t\right)^2\right]\right\},
   \label{eqn:Entropy}
\end{align}
where $c$ is an $\mathcal{O}(10^{-1})$ parameter defined in Appendix \ref{app:phase entropy}. Comparing  the entropy (\ref{eqn:covent}) arising for  the individual initial two-particle product state components (\ref{initeq}) with the relative phase entanglement entropy (\ref{eqn:Entropy}), we see that the former scales as $\alpha^{-6}$ while the latter scales as $\alpha^{-2}$. Recall that $\alpha$ is chosen as a real number that can be interpreted as the initial maximum displacement of the coherent state position space wave function in units of the harmonic oscillator zero point uncertainty. In any conceivable, realistic implementation, we expect $\alpha\ggg 1$, so that the entanglement entropy for the relative phase terms dominates over the entropy arising for the single component terms. This justifies the approximation of keeping only the relative phase terms in the time-evolved state for the purposes of determining the gravitationally-induced entanglement between the two harmonically trapped particles.

\begin{figure}
    \centering
    \includegraphics[scale=0.5]{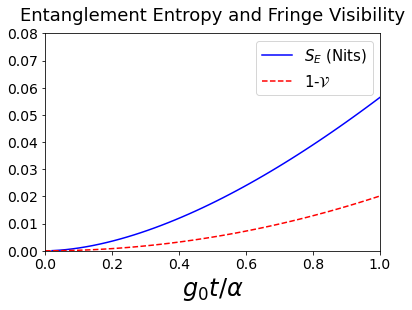}
    \caption{Entanglement entropy $S$ (units of nits, where 1 nit $=\frac{1}{\ln 2}$ bits) for the reduced state of one of the particles and the fringe visibility $1-{\mathcal{V}}$ for particle 1 as a function of the scaled time parameter $\frac{g_0}{\alpha}t$.} 
    \label{fig:EntropyandVisibility}
\end{figure}

While the growth of entanglement entropy as given by Eq. (\ref{eqn:Entropy}) demonstrates that the evolving  state [resulting from the initial  state (\ref{initeq})] develops genuine entanglement between the two particles, the entanglement entropy is not itself an observable; we still require a means for measuring the entanglement. Consider the probability distribution for detecting particle 1 on the $x$ axis. In terms of the approximate evolved state (\ref{approxstateq}), we have
\begin{equation}
\label{eqn:probDist}
\begin{split}
    \left|\bra{x_1}\ket{\psi}\right|^2=&|N|^2\,\left\{ 2\left[1+\mathrm{Re}\left(e^{i\left(\delta S_{aa}-\delta S_{ab}\right)}\bra{\bm{\alpha}_{2b}}\ket{\bm{\alpha}_{2a}}\right)\right]|\psi_{1a}\left(x_1\right)|^2\right.\\
   &\left.+2\left[1+\mathrm{Re}\left(e^{i\left(\delta S_{ba}-\delta S_{bb}\right)}\bra{\bm{\alpha}_{2b}}\ket{\bm{\alpha}_{2a}}\right)\right]|\psi_{1b}\left(x_1\right)|^2\right.\\
   &\left.+2\mathrm{Re}\left[\psi_{1a}\left(x_1\right)\psi^{*}_{1b}\left(x_1\right)
   \left(e^{i\left(\delta S_{aa}-\delta S_{ba}\right)}+e^{i\left(\delta S_{ab}-\delta S_{bb}\right)}\right.\right.\right.\\
 &\left.\left.\left.+e^{i\left(\delta S_{aa}-\delta S_{bb}\right)}\bra{\bm{\alpha}_{2b}}\ket{\bm{\alpha}_{2a}}+e^{i\left(\delta S_{ab}-\delta S_{ba}\right)}\bra{\bm{\alpha}_{2a}}\ket{\bm{\alpha}_{2b}}\right)\right]\right\},
\end{split}
\end{equation}
where $\psi_{1a}\left(x_1\right)=\bra{x_1}\ket{\bm{\alpha}_{1a}}$ etc. When the average positions for $\psi_{1a}$ and $\psi_{1b}$ overlap, which corresponds to their classical paths crossing in configuration space, we have $|\psi_{1a}|=|\psi_{1b}|$. The third term in the above expression will then have spatially oscillatory terms:
\begin{equation}
\begin{split}
    \psi_{1a}\psi_{1b}^*=&
    |\psi_{1a}|^2\exp\left[ i\sqrt{2}\left(\mathrm{Im}\,\alpha_{1ax}-\mathrm{Im}\,\alpha_{1bx}\right)\left(x_1-\frac{1}{\sqrt{2}}\mathrm{Re}\,\alpha_{1ax}\right)\right]=|\psi_{1a}|^2e^{i\Phi(x_1)},
\end{split}
\end{equation}
where here $\Phi$  denotes the phase term in the middle expression. 
These oscillatory terms appear as fringes modulated by a Gaussian envelope, as indicated schematically in Fig. \ref{fig:setup}. Expressing the probability distribution (\ref{eqn:probDist}) in the form $|\bra{x_1}\ket{\psi}|^2=\chi+2[\mathrm{Re}\,\mu\cos\Phi(x_1)-\mathrm{Im}\,\mu\sin\Phi(x_1)]$, 
the extrema of the distribution occur at $\chi\pm2|\mu|$. The fringe visibility $\mathcal{V}$,  defined as the ratio of the difference between the extrema to their sum, is given approximately as
\begin{equation}
    \mathcal{V}=\frac{2|\mu|}{\eta}\approx1-\frac{1}{8}\left(\frac{c g_0 t}{\alpha}\right)^2
    \label{visibility}
\end{equation}
for $\alpha\gg 1$;
the constant $c$ is the same as that appearing in  Eq. (\ref{eqn:Entropy}). Comparing Eq. (\ref{eqn:Entropy}) with Eq. (\ref{visibility}), we can see that the visibility scales monotonically with the entanglement entropy (Fig. \ref{fig:EntropyandVisibility}).  As the entanglement entropy increases, the reduced state for either particle behaves increasingly as an effective classical mixture of coherent states and so the quantum interference in the particle position detection probability distribution decreases correspondingly. Therefore, the  fringe visibility can be employed as an operational measure of the entanglement entropy for probing gravitationally induced entanglement in our harmonic particle trap scheme. Using, for example, masses $m\sim 10^{-14}\, \mathrm{kg}$ and a separation distance  $\sim 10^2\,\mu \mathrm{m}$ as in Ref. \cite{bose2017}, and an optical trapping potential frequency $\omega\sim10^2\, \mathrm{kHz}$ as in Ref. \cite{delic2020}, gives a value $\frac{g_0}{\alpha}\sim 10^{-6}$ which would result in percentage level changes in the fringe visibility in times of order one second.

Recently, a rather general quantum interferometric identity was obtained that relates the concurrence ($\mathcal{C}$), visibility ($\mathcal{V}$), and which-path information (``particleness" $\mathcal{P}$) as complementary measures for certifying the quantum aspects of gravitational entanglement interactions \cite{Maleki_2022}:  
\begin{equation}
 \mathcal{V}^2+\mathcal{P}^2+\mathcal{C}^2=1.
 \label{malekieq}
\end{equation}
 For the entangled state considered in this work, and in the limit of large $\alpha$ where we can treat the coherent state component pairs as mutually orthogonal to a good approximation, our state can be alternatively interpreted as that arising for the two interferometer setup described in Ref. \cite{Maleki_2022}, and using the forms for $\mathcal{P,V,}$ and $\mathcal{C}$ given there. For our system state, we have: $\mathcal{P}\approx|(|e^{i\delta S_{aa}}|^2+|e^{i\delta S_{ab}}|^2)-(|e^{i\delta S_{ba}}|^2+|e^{i\delta S_{bb}}|^2)|=0$, $\mathcal{V}^2\approx 2|e^{i\delta S_{aa}-i\delta S_{ba}}+e^{i\delta S_{aa}-i\delta S_{bb}}|\approx 1-\frac{1}{4}\left(\frac{cg_0}{\alpha}t\right)^2$ and $\mathcal{C}^2\approx 2|e^{i\delta S_{aa}+i\delta S_{bb}}-e^{i\delta S_{ab}+i\delta S_{ba}}|\approx\frac{1}{4}\left(\frac{cg_0}{\alpha}t\right)^2$, hence obeying the  identity (\ref{malekieq}) to the leading order in $\alpha^{-1}$ with which we calculate these measures.

\section{Alternative Potentials}
\label{sec:potentials}
The method of approximating the evolution of the state considered in this work is not unique to the Newtonian gravitational potential; it can also be applied to other weak interactions that are treated as perturbations to the harmonic trap. Of potential interest is a Coulomb type potential generalized to arbitrary dimensions or a Yukawa potential \cite{barker_entanglement_2022}. Coulomb potentials of various dimensions may be useful if practical applications of this proposal use extended objects to effectively reduce the dimension  of the interaction. Yukawa potentials arise in theories with massive gravitons or in theories containing light scalar particles such as dilatons, and both could be used as alternatives to test for possible deviations from the inverse square law \cite{mostepanenko_state_2020}. Here, we calculate the effect these different potentials would have on the entanglement entropy or fringe visibility.

\begin{figure}
    \centering
    \begin{subfigure}[b]{0.45\textwidth}
        \centering
        \includegraphics[scale=0.6]{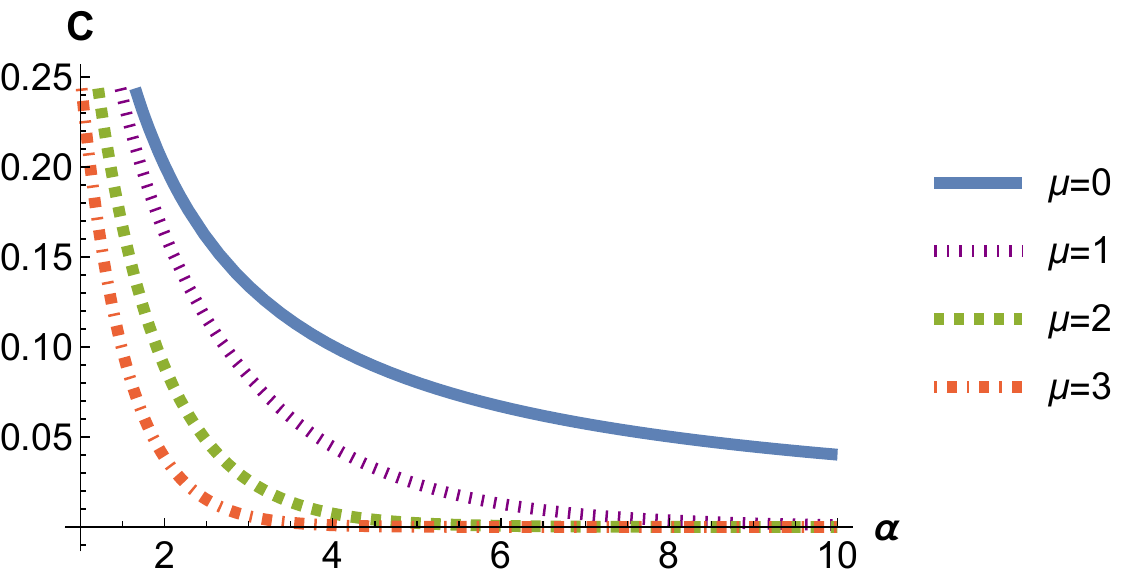}
        \caption{Yukawa}
    \end{subfigure}
    \begin{subfigure}[b]{0.45\textwidth}
        \centering
        \includegraphics[scale=0.6]{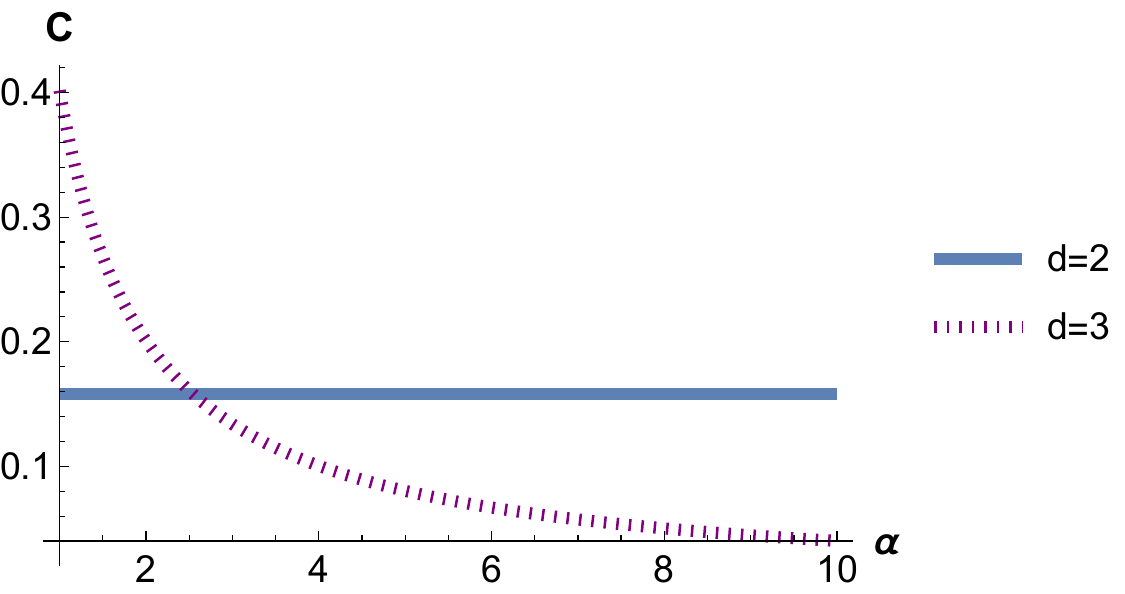}
        \caption{Coulomb}
    \end{subfigure}
    \caption{Value of the constant $C$ given in the text which controls the scaling of entanglement entropy and fringe visibility for (a) Yukawa potential with various values of the Yukawa mass $\mu$, and (b) Coulomb potential with dimension, $d=2,3$, as a function of the initial coherent state eigenvalue $\alpha$, with $g_0=1$.}
    \label{fig:Yukawa}
\end{figure}

Changing the potential results in different values for the first order corrections to the classical trajectories used in the calculation of the entanglement entropy given by Eq. (\ref{eqn:Entropy}), 
which in turn results in different forms for the $\delta S$ terms that appear in the phase factors of the time evolved state. Since the entanglement entropy and fringe visibility can be expressed as functions of these $\delta S$ terms, it suffices to compute the modified forms of the latter. A new interaction potential equates to changing the Kameltonian $K$ in Eq. (\ref{kameltonian}), which will result in a modification to the values
\begin{align}
    \dot{\bar{\eta}}_{1x}&=\frac{1}{2\pi}\int_0^{2\pi}-\frac{\partial K}{\partial \phi_x}\Big|_0dt=\frac{1}{2\pi}\int_0^{2\pi}-\frac{\partial K}{\partial r}\frac{\eta_{x0}\sin\left(t+\phi_{x0}\right)\cos\left(t+\phi_{x0}\right)}{r}\Big|_0dt\\
    \dot{\bar{\phi}}_{1x}&=\frac{1}{2\pi}\int_0^{2\pi}\frac{\partial K}{\partial \eta_x}\Big|_0dt=\frac{1}{2\pi}\int_0^{2\pi}\frac{\partial K}{\partial r}\frac{\sin^2\left(t+\phi_{x0}\right)}{r}\Big|_0dt,
\end{align}
and similarly for the $y$ components. Here, $\eta$ and $\phi$ are the Hamilton Jacobi coordinates for the classical harmonic oscillator which are used in our approximation in Appendix \ref{app:heller apx}. For the initial conditions considered above, i.e., $\alpha_{1ax}=\alpha,\alpha_{1bx}=\alpha e^{-i\pi/4},\alpha_{2ay}=\alpha e^{i\pi/2},$ and $\alpha_{2by}=\alpha e^{i\pi/4}$, these modifications result in phase terms given by
\begin{align}
    \delta S_{aa}=\delta S_{bb}&=\left(\frac{1}{2}\alpha\frac{\partial K}{\partial r}-K\right)\Big|_{r=\alpha}t,\\
    \delta S_{ab}+\delta S_{ba}&=2\left(\alpha^2\dot{\bar{\phi}}_{1x}-\bar{K}\right)\Big|_{r=r_{ab}}t.
\end{align}
Here, $\delta S_{aa}$ and $\delta S_{bb}$ are evaluated along $r=\alpha$ which is a constant in the first order classical canonical perturbation theory approximation. The terms $\delta S_{ab}$ and $\delta S_{ba}$ can be evaluated along the trajectory for $r$ in the \textit{ab} component and we only use the perturbation to $\phi_x$, and not $\phi_y$, due to the symmetries of the initial conditions. The expressions for the entanglement entropy and fringe visibility can be recovered from Eqs. (\ref{eqn:Entropy}) and (\ref{visibility}), but now with the constant $\frac{c g_0}{\alpha}$ replaced by
\begin{equation}
    C=2\left(\alpha^2 \dot{\bar{\phi}}_{1x}-\bar{K}\right)\Big|_{r=r_{ab}}+\left(2K-\alpha\frac{\partial K}{\partial r} \right)\Big|_{r=\alpha}.
\end{equation}
We calculate this quantity for a Coulomb potential in arbitrary dimension, i.e.
\begin{equation}
    K=
    \begin{cases}
    -g_0\,r^{2-d} & d> 2\\
    g_0 \log{r} & d=2
    \end{cases},
\end{equation}
and for a Yukawa potential with
\begin{equation}
    K=g_0 \frac{e^{-\mu r}}{r},
\end{equation}
where $\mu$ is related to the mass of the mediating particle. The values of $C$ for the Yukawa potential with various values of the mass $\mu$ and for the Coulomb potential in various dimensions $d$ are shown in Fig. \ref{fig:Yukawa}. These different values of the parameter $C$ would result in different scalings for the entanglement entropy and fringe visibility. With precise enough visibility measurements, such differences could serve as a probe for possible new quantum modified gravity theories through their resulting particle entanglement.

\section{Conclusion}{\label{sec:conclusion}}
Merging quantum mechanics and gravity is one of the loftiest goals of modern physics. Recent work investigating gravitationally induced quantum entanglement provides a pathway to new insights into this problem.

In this paper, we have developed a thought experiment model to detect gravitationally induced entanglement at the level of non-relativistic quantum mechanics, similar to other proposals in the literature. Advantageously, our setup  readily lends itself to a more fundamental quantum field theoretic description, to be presented  in a subsequent work, wherein the massive particles which become entangled through gravity are excitations of a quantum scalar field, and the fringe visibility is modeled by an expectation value of field observables.

This quantum field theoretic approach will allow us to go beyond the predictions of our non-relativistic quantum mechanical model in several ways. First of all, we will be able to systematically account for  relativistic corrections to our predicted measure of entanglement, corresponding to the limit of a large harmonic oscillator trapping potential frequency. Secondly, since different methods of carrying out this approach, such as the closed time path formalism \cite{campos1998,calzetta2008}, or the Dirac constraint quantization procedure \cite{oniga2016,oniga2017}, treat gauge fixing in different ways, we may be able to better understand more comprehensively the gauge  aspects of gravitational entanglement generation. Apart from these insights, there may be other, less expected, lessons to be learned. As mentioned in the introduction, the description of gravitational entanglement in terms of non-relativistic quantum mechanics is not sensitive to the spin (i.e., helicity) of the mediating massless boson \cite{Carney2022}; a  relativistic, quantum field theoretic approach that accounts for retardation effects, such as the present proposed approach, is required in order to reveal the distinguishing helicity two, graviton nature of the entanglement.

\section*{Acknowledgements}
We thank Charis Anastopoulos, Markus Aspelmeyer, Bei-Lok Hu, David Mattingly, Sougato Bose, Alexander Smith, Qidong Xu, and Shadi Ali Ahmad for very helpful conversations This work was supported by the NSF under Grant No. PHY-2011382.

\bibliography{apssamp}

\appendix
\section{Approximate Evolution of Gaussian Parameters}
\label{app:heller apx}
In order to find the solutions to Eqs. (\ref{eqn:Heller1})-(\ref{eqn:Heller3}), restated here,
\begin{align}
    \expval{\dot{\bm{r}}}&=\frac{\partial H}{\partial \expval{\bm{p}_r}},\quad\expval{\dot{\bm{p}_r}}=-\frac{\partial H}{\partial \expval{\bm{r}}},\label{eqn:Heller1A}\\
    \dot{\Omega}&=-2\Omega^2-\frac{1}{2}\frac{\partial^2 V}{\partial r_i \partial r_j}\Big|_{\bm{r}=\expval{\bm{r}}},\label{eqn:Heller2A}\\
    \dot{\gamma}&=i\, \mathrm{Tr}\,\Omega+\mathcal{L}.
    \label{eqn:Heller3A}
\end{align}
we introduce a bookkeeping parameter $\lambda$ into the Hamiltonian (\ref{scomhameq}), multiplying the gravitational potential term
\begin{equation}
        H=\frac{1}{2}\left(p_r^2 +r^2-\lambda\frac{g_0}{r}\right).
        \label{hamwLambdaeq}
\end{equation}
Thus we treat the gravitational term as a perturbation to the harmonic oscillator, and at the end of the calculation take $\lambda \rightarrow1$; this procedure is validated {\it a posteriori} with the relative magnitudes of the terms in the resulting perturbation series being sufficiently small. For $\expval{\bm{r}}$ and $\expval{\bm{p}_r}$, we use classical canonical perturbation theory, moving to the Hamilton-Jacobi coordinates of the 2D harmonic oscillator given by
\begin{equation}
    \expval{r_x}=\sqrt{2\eta_x}\sin\left(t+\phi_x\right),\ \expval{r_y}=\sqrt{2\eta_y}\sin\left(t+\phi_y\right).
\end{equation}
The Hamilton-Jacobi coordinates are those generated by a canonical transformation where the unperturbed Hamiltonian is zero and therefore these coordinates are constants. However, the gravitational term remains and so the transformed Hamiltonian (Kameltonian)  in these coordinates is given by 
\begin{equation}
    K=-\lambda\frac{g_0}{\sqrt{2}}[\eta_x\sin^2\left(t+\phi_x\right)+\eta_y\sin^2\left(t+\phi_y\right)]^{-1/2}.
    \label{kameltonian}
\end{equation}
We assume a perturbative solution and to first order the equations of motion are given by
\begin{align}
\dot{\eta}_{x1}&=-\frac{\partial K}{\partial \phi_x}\Big|_0=-\lambda g_0\frac{\eta_{x0}\sin\left(t+\phi_{x0}\right)\cos\left(t+\phi_{x0}\right)}{\sqrt{2}[\eta_{x0}\sin^2\left(t+\phi_{x0}\right)+\eta_{y0}\sin^2\left(t+\phi_{y0}\right)]^{3/2}},\\
\dot{\phi}_{x1}&=\frac{\partial K}{\partial \eta_x}\Big|_0=\lambda g_0\frac{\sin^2\left(t+\phi_{x0}\right)}{2\sqrt{2}[\eta_{x0}\sin^2\left(t+\phi_{x0}\right)+\eta_{y0}\sin^2\left(t+\phi_{y0}\right)]^{3/2}},
\end{align}
and similarly for the $y$ components.
The right hand side of these equations is periodic and we approximate the solutions by their time averages over an oscillator period as follows:
\begin{align}
\eta_{x1}&\approx\dot{\overline{\eta}}_{x1}t=-\lambda\frac{g_0 t}{2\sqrt{2}\pi}\int_0^{2\pi}\frac{\eta_{x0}\sin\left(t+\phi_{x0}\right)\cos\left(t+\phi_{x0}\right)}{[\eta_{x0}\sin^2\left(t+\phi_{x0}\right)+\eta_{y0}\sin^2\left(t+\phi_{y0}\right)]^{3/2}}dt,\\
\phi_{x1}&\approx\dot{\overline{\phi}}_{x1}t=\lambda\frac{g_0 t}{4\sqrt{2}\pi}\int_0^{2\pi}\frac{\sin\left(t+\phi_{x0}\right)^2}{[\eta_{x0}\sin^2\left(t+\phi_{x0}\right)+\eta_{y0}\sin^2\left(t+\phi_{y0}\right)]^{3/2}}dt,
\end{align} 
and similarly for the $y$ components. For initial conditions given by $\alpha_1=(\alpha,0)$ and $\alpha_2=(0,\alpha e^{i\pi/2})$ for some $\alpha>0$, the relative difference coordinate coherent state parameters are $\alpha_{rx}=\frac{\alpha}{\sqrt{2}}$ and $\alpha_{ry}=-i\frac{\alpha}{\sqrt{2}}$. This leads to the following first order solutions:
\begin{equation}
    \expval{r_x}=\alpha \cos\left[\left(1+\lambda\frac{g_0}{2\alpha^3}\right) t\right],\ \expval{r_y}=-\alpha\sin \left[\left(1+\lambda\frac{g_0}{2\alpha^3}\right) t\right].
\end{equation}
Such a time average has the advantage that $\expval{r}=\alpha$ is  constant. 

We can only approximate $\Omega$ analytically  for these particular initial conditions where $\expval{r}=\alpha$ is  constant. Assuming a solution of the form $\Omega\approx\Omega_0+\lambda\Omega_1$ to Eq. (\ref{eqn:Heller2}) for the initial conditions $\alpha_1=(\alpha,0)$ and $\alpha_2=(0,\alpha e^{i\pi/2})$ this results in the following equations of motion:
\begin{flalign}
    \dot{\Omega}_0=&-2\Omega_0^2-\mathbb{1},\\
    \dot{\Omega}_1=&-2(\Omega_0\Omega_1 +\Omega_1\Omega_0) \nonumber\\
    &-\frac{g_0}{2\alpha^3}\begin{bmatrix}        1-3\cos^2\left[\left(1+\lambda\frac{g_0}{2\alpha^3}\right) t\right]&&3\cos\left[\left(1+\lambda\frac{g_0}{2\alpha^3}\right) t\right]\sin\left[\left(1+\lambda\frac{g_0}{2\alpha^3}\right) t\right]\\
    3\cos\left[\left(1+\lambda\frac{g_0}{2\alpha^3}\right) t\right]\sin\left[\left(1+\lambda\frac{g_0}{2\alpha^3}\right) t\right]&&1-3\sin^2\left[\left(1+\lambda\frac{g_0}{2\alpha^3}\right) t\right]        
    \end{bmatrix}.
\end{flalign}
 The initial condition for all coherent state parameters is given by $\Omega=\frac{i}{2}\mathbb{1}$. This gives a solution to the equation for $\Omega_1$, which we then expand as a  series in $\lambda$, but keep terms where it appears in the phases of oscillating terms. We find that the solution has a lowest order term proportional to $\lambda^{-1}$. This is due to the fact that we are taking an `improper' series expansion by leaving in the phase terms. However if we assume an approximate solution for $\Omega$ given by $\Omega_0$ and the $\lambda^{-1}$ order term of $\Omega_1$ we find that this is also a solution of the zeroth order equation of motion:
\begin{equation}
    \Omega=\frac{i}{2}\mathbb{1}+\frac{3}{8}\left[e^{-i\left(2+\frac{g_0}{\alpha^3}\right)t}-e^{-2it}\right]
    \begin{bmatrix}
    i&&1\\
    1&&-i
    \end{bmatrix},
\end{equation}
where we have taken $\lambda\rightarrow1$ and note that this expression is still valid for $\alpha\gg1$. The off-diagonal terms clearly make the state nonseparable in the $r_x$ and $r_y$ coordinates, and upon transforming back to the particle coordinates, the scaled covariance matrix which we will denote $\Omega_{12}$ is given by
 \begin{equation}
     \Omega_{12}=\frac{i}{2}\mathbb{1}+\frac{3}{16}\left[e^{-i\left(2+\lambda\frac{ g_0}{\alpha^3}\right)t}-e^{-2it}\right]
     \begin{bmatrix}
    i&&1&&i&&-1\\
    1&&-i&&-1&&i\\
    i&&-1&&i&&1\\
    -1&&i&&1&&-i
     \end{bmatrix},
 \end{equation}
 which implies that this is an entangled state between particles 1 and 2.
 
 Thus, gravity gives rise to entanglement without considering a superposition of coherent state products. In order to quantify this entanglement, we evaluate the entanglement entropy using the formalism for Gaussian states \cite{demarie_pedagogical_2012}. The Wigner function for this state is given by a Gaussian distribution on phase space whose covariance matrix, in the two particle phase space coordinates basis, is given by $\left(\bm{x}_1,\bm{x}_2,\bm{p}_1,\bm{p}_2\right)$, is
\begin{equation}
    \Sigma=
    \begin{bmatrix}
    \frac{1}{2}\left(\mathrm{Im}\,\Omega\right)^{-1}&&-\left(\mathrm{Im}\,\Omega\right)^{-1}\mathrm{Re}\,\Omega\\
    Re\,\Omega\left(\mathrm{Im}\,\Omega\right)^{-1}&&2\left[\mathrm{Im}\,\Omega+\mathrm{Re}\,\Omega\left(\mathrm{Im}\,\Omega\right)^{-1}\mathrm{Re}\,\Omega\right]
    \end{bmatrix}.
\end{equation}
The symplectic eigenvalues, $\nu_i$, of one of the subsystems are obtained by determining the eigenvalues of $iJ\Sigma_1$, where $\Sigma_1$ is the submatrix of the covariance matrix corresponding to particle 1 and $J$ is the symplectic form given by 
\begin{equation}
    J=
    \begin{bmatrix}
    0&&\mathbb{1}\\
    -\mathbb{1}&&0
    \end{bmatrix}.
\end{equation}
The entanglement entropy is then given by
\begin{align}
    S=&\sum_i \frac{1}{2}\left(\nu_i+1\right)\log\left[\frac{1}{2}\left(\nu_i+1\right)\right]-\frac{1}{2}\left(\nu_i-1\right)\log\left[\frac{1}{2}\left(\nu_i-1\right)\right]\nonumber\\
    &\approx\frac{9 g_0^2 t^2 }{64 \alpha^6}\left[1-\log \left(\frac{9 g_0^2 t^2}{128 \alpha^6}\right)\right],
\end{align}
where we sum over the symplectic eigenvalues for one of the particles. The bottom expression here contains the lowest order terms in $\alpha^{-1}$.

\section{Approximate Phase Dependent Entanglement Entropies}
\label{app:phase entropy}
For initial conditions given by $\alpha_{1ax}=\alpha,\alpha_{1bx}=\alpha e^{-i\pi/4},\alpha_{2ay}=\alpha e^{i\pi/2},$ and $\alpha_{2by}=\alpha e^{i\pi/4}$, we find the terms $\delta S$ for the various components by using the classical canonical perturbation theory method described in appendix \ref{app:heller apx} to solve for the classical paths. We then use these trajectories to calculate the difference in action between the perturbed and unperturbed (pure harmonic oscillator) solutions. We also take $\lambda\rightarrow1$ here and find that our approximation are still valid for $\alpha\gg1$. The $\delta S$ terms are given by
\begin{align}
\delta S_{aa}=&\delta S_{bb}= \frac{3g_0}{2\alpha}t\\
\delta S_{ab}=&\frac{\alpha^2}{4}\left\{\cos\left[2\left(1+C_{\phi}\frac{g_0}{\alpha^3}\right)t\right]-\sin\left[2\left(1+C_{\phi}\frac{g_0}{\alpha^3}\right)t\right]-\cos\left(2t\right)+\sin\left(2t\right)\right\}\nonumber\\
&+\frac{g_0}{\alpha}t\left(C_{\phi}+\frac{1}{2}C_{\lambda}\left\{\cos\left[2\left(1+C_{\phi}\frac{g_0}{\alpha^3}\right)t\right]+\sin\left[2\left(1+C_{\phi}\frac{g_0}{\alpha^3}\right)t\right]\right\}+C_{gab}\right)\\
\delta S_{ba}=&
-\frac{\alpha^2}{4}\left\{\cos\left[2\left(1+C_{\phi}\frac{g_0}{\alpha^3}\right)t\right]-\sin\left[2\left(1+C_{\phi}\frac{g_0}{\alpha^3}\right)t\right]-\cos\left(2t\right)+\sin\left(2t\right)\right\}\nonumber\\
&+\frac{g_0}{\alpha}t\left(C_{\phi}-\frac{1}{2}C_{\lambda}\left\{\cos\left[2\left(1+C_{\phi}\frac{g_0}{\alpha^3}\right)t\right]+\sin\left[2\left(1+C_{\phi}\frac{g_0}{\alpha^3}\right)t\right]\right\}+C_{gba}\right),
\end{align}
where $C_{\phi},C_{\lambda},C_{gab},$ and $C_{gba}$ are given respectively by

\begin{align}
C_{\phi}=&\frac{1}{2\pi}\int_0^{2\pi}dt\frac{\cos ^2\left(t+\frac{\pi }{4}\right)}{\left(\sin ^2(t)+\cos ^2\left(t+\frac{\pi }{4}\right)\right)^{3/2}}\\
C_{\lambda}=&\frac{1}{2\pi}\int_0^{2\pi}dt\frac{\sin (t) \cos (t)}{\left(\sin ^2(t)+\cos ^2\left(t+\frac{\pi }{4}\right)\right)^{3/2}}\\
C_{gab}=&\frac{1+C_{\phi}\frac{g_0}{\alpha^3}}{\pi}\int_0^{\frac{\pi}{1+C_{\phi}\frac{g_0}{\alpha^3}}}dt\left(\sin ^2\left(t+\frac{\pi }{4}\right)+\cos ^2(t)\right)^{-1/2}\\
C_{gba}=&\frac{1+C_{\phi}\frac{g_0}{\alpha^3}}{\pi}\int_0^{\frac{\pi}{1+C_{\phi}\frac{g_0}{\alpha^3}}}dt\left(\sin ^2(t)+\cos ^2\left(t+\frac{\pi }{4}\right)\right)^{-1/2}.
\end{align}

The reduced density matrix for particle 1 is given by
\begin{equation}
    \begin{split}
        \rho_1=&|N|^2\left\{\ket{\bm{\alpha}_{1a}}\bra{\bm{\alpha}_{1a}}\left[2+2e^{\alpha^2\left(\frac{1}{\sqrt{2}}-1\right)}\cos\left(\delta S_{aa}-\delta S_{ab}+\frac{\alpha^2}{\sqrt{2}}\right)\right]\right.\\
        &+\ket{\bm{\alpha}_{1a}}\bra{\bm{\alpha}_{1b}}\left[e^{i\left(\delta S_{ab}-\delta S_{bb}\right)}+e^{i\left(\delta S_{aa}-\delta S_{ba}\right)}\right.\\
        &\left.+e^{\alpha^2\left(\frac{1}{\sqrt{2}}-1\right)}\left(e^{i\left(\delta S_{aa}-\delta S_{bb}+\frac{\alpha^2}{\sqrt{2}}\right)}+e^{i\left(\delta S_{ab}-\delta S_{ba}-\frac{\alpha^2}{\sqrt{2}}\right)}\right)\right]\\
        &+\ket{\bm{\alpha}_{1b}}\bra{\bm{\alpha}_{1a}}\left[e^{-i\left(\delta S_{ab}-\delta S_{bb}\right)}+e^{-i\left(\delta S_{aa}-\delta S_{ba}\right)}\right.\\
        &\left.+e^{\alpha^2\left(\frac{1}{\sqrt{2}}-1\right)}\left(e^{-i\left(\delta S_{aa}-\delta S_{bb}+\frac{\alpha^2}{\sqrt{2}}\right)}+e^{-i\left(\delta S_{ab}-\delta S_{ba}-\frac{\alpha^2}{\sqrt{2}}\right)}\right)\right]\\
        &\left.\ket{\bm{\alpha}_{1b}}\bra{\bm{\alpha}_{1b}}\left[2+2e^{\alpha^2\left(\frac{1}{\sqrt{2}}-1\right)}\cos\left(\delta S_{ba}-\delta S_{bb}+\frac{\alpha^2}{\sqrt{2}}\right)\right]\right\}.
    \end{split}
\end{equation}
We shall denote the four outer product terms in the curly brackets of this expression respectively as $\rho_{aa}$, $\rho_{ab}$, $\rho_{ba}$, and $\rho_{bb}$. 
The eigenvalues of the density matrix $\sigma_i$ are given by
\begin{equation}
\begin{split}
    \sigma_i=&\frac{|N|^2}{2}\left[\rho_{aa}+\rho_{bb}+2\mathrm{Re}\,\left(e^{\alpha^2\left(e^{i\pi/4}-1\right)}\rho_{ab}\right)\pm\left(\left(\rho_{aa}-\rho_{bb}\right)^2+4\left\{e^{\alpha^2(\sqrt{2}-2)}\rho_{aa}\rho_{bb}+|\rho_{ab}|^2\right.\right.\right.\\
    &\left.\left.\left.+\left(\rho_{aa}+\rho_{bb}\right)\mathrm{Re}\,\left(e^{\alpha^2(e^{i\pi/4}-1)}\rho_{ba}\right)-\left[\mathrm{Im}\,\left(e^{\alpha^2(e^{i\pi/4}-1)}\rho_{ba}\right)\right]^2\right\}\right)^{1/2}\right],
\end{split}
\end{equation}
where $i=\{1,2\}$. The coherent state components of particle 1 are orthogonal to a good approximation for $\alpha\gg1$. In this limit, the reduced state for particle 1 can be written as
\begin{equation}
\begin{split}
      \rho_1\approx&\frac{1}{4} \{2\ket{\bm{\alpha}_{1a}}\bra{\bm{\alpha}_{1a}}+\left(e^{i(\delta S_{ab}-\delta S_{bb})}+e^{i(\delta S_{aa}-\delta S_{ba})}\right)\ket{\bm{\alpha}_{1a}}\bra{\bm{\alpha}_{1b}}\\
      &+\left(e^{-i\left(\delta S_{ab}-\delta S_{bb}\right)}+e^{-i\left(\delta S_{aa}-\delta S_{ba}\right)}\right)\ket{\bm{\alpha}_{1b}}\bra{\bm{\alpha}_{1a}}+2\ket{\bm{\alpha}_{1b}}\bra{\bm{\alpha}_{1b}} \} ,  
\end{split}
\end{equation}
and the eigenvalues can be approximated as
 \begin{equation}
     \sigma_i\approx\frac{1}{2}\pm\frac{1}{4}|\rho_{ab}|,
 \end{equation}
 with
 \begin{equation}
     \begin{split}
        |\rho_{ab}|\approx&\left[2+2\cos\left(\delta S_{ab}-\delta S_{bb}-\delta S_{aa}+\delta S_{ba}\right)\right]^{1/2}=\left\{2+2\cos\left[\left(2C_{\phi}+C_{gab}+C_{gba}-3\right)\left(\frac{g_0 t}{\alpha}\right)\right]\right\}^{1/2}\\
        \approx&2-\frac{1}{4}\left(2C_{\phi}+C_{gab}+C_{gba}-3\right)^2\left(\frac{g_0 t}{\alpha}\right)^2.
     \end{split}
 \end{equation}
The entanglement entropy is then approximately
 \begin{equation}
    \begin{split}
        S=&-\sigma_1\log\sigma_1-\sigma_2\log\sigma_2\\
        \approx&\frac{1}{16}\left(2C_{\phi}+C_{gab}+C_{gba}-3\right)^2\left(\frac{g_0 t}{\alpha}\right)^2\\
        &\left\{1-\log\left[\frac{1}{16}\left(2C_{\phi}+C_{gab}+C_{gba}-3\right)^2\left(\frac{g_0 t}{\alpha}\right)^2\right]\right\} .  
    \end{split}
\end{equation}
 
This entanglement grows faster than that due to the covariance terms in one component of our full state by a factor of $\alpha^{4}$, which will be very large in any realistic implementation, so our approximation where the local entanglement terms are ignored is justified. Noting this, we can further simplify the above expression by approximating $C_{gab}$, and $C_{gba}$ to find that they only have corrections proportional to $\alpha^{-3}$ which are negligible. Therefore, we have

\begin{align}
    C_{gab}\approx&\,C_{gba}\approx\frac{1}{2\pi}\int_0^{2\pi}dt\left\{\sin^2\left(t+\frac{\pi}{4}\right)+\cos^2t\right\}^{-1/2},
\end{align}
giving the following expression for the entanglement entropy:
\begin{align}
   S\approx&\left(\frac{cg_0 }{\alpha}t\right)^2\left\{1-\log\left[\left(\frac{cg_0 }{\alpha}t\right)^2\right]\right\}, 
\end{align}
where $c=\frac{1}{4}\left(2C_{\phi}+2C_{gab}-3\right)\approx0.1004$.

\section{Numerical Validation of Analytical Approximation to Time Evolution}
\label{app:numerics}
In order to validate the analytical approximation to the time evolved state of the gravitational harmonium described above, we also numerically approximate the time evolution of this state and compare the numerical and analytical approximations. In order to do this we choose a basis for our state space given by simultaneous eigenstates of the two dimensional harmonic oscillator Hamiltonian and the angular momentum operator, whose position space wave functions are given by
\begin{equation}
    \bra{r,\phi}\ket{n,\pm l}=\left(\frac{n!}{\pi\left(n+l\right)!}\right)^{1/2}e^{\pm il\phi}e^{-r^2/2}r^lL_n^l(r^2)
\end{equation}
where $l>0$, and $L_n^l$ are the generalized Laguerre polynomials \cite{COLAVITA2005183}. The energy of these states is $E_{n,l}=2n+l+1$ for a simple harmonic oscillator, in the absence of gravity.

The gravitational potential term can be evaluated in this basis with matrix elements given by
\begin{align}
    V_{nmjl}=&\bra{n,l}-\frac{g_0}{r}\ket{m,j}=-2g_0\delta_{jl}\left(\frac{n!m!}{(n+l)!(m+j)!}\right)^{1/2}\int_0^{\infty}e^{-r^2}r^{j+l}L_n^l(r^2)L_m^j(r^2)dr\\
    =&-g_0\sum_{i=0}^N\frac{\Gamma(N+l+1/2)x_i}{N!(N+1)^2(L_{N+1}^{l-1/2}(x_i))^2}L_n^l(x_i)L_m^l(x_i),\nonumber
\end{align}
where $N>n,m$, the $x_i$ are the zeros of $L_N^{l-1/2}$, and the last equality is found by Gauss-Laguerre quadrature. Therefore the Hamiltonian for the gravitational harmonium has matrix elements in this basis given by $H_{nmlj}=\left(2n+l+1\right)\delta_{nm}\delta_{jl}+V_{nmjl}$. By evaluating a finite number of these basis elements, we approximate the Hamiltonian on the subspace spanned by the finite number of basis states corresponding to these matrix elements. We then numerically diagonalize this matrix and approximate the propagator on this subspace by $U(t)=Pe^{-iEt}P^{\dagger}$, where $P$ is the change of basis matrix to the eigenbasis of $H$, and $E$ is the diagonal matrix of eigenvalues of $H$.

We act with this propagator on an initial state corresponding to one component of our initial state considered above, i.e., a product of coherent states of particles 1 and 2. This corresponds to a coherent state of a 2D harmonic oscillator in the separation coordinate we are considering here. In this basis, coherent states are given by
\begin{equation}
    \ket{\alpha,\beta}_{\pm}=\sum_{n,l=0}^{\infty}\left(\frac{(n+l)!}{n!}\right)^{1/2}\frac{1}{l!}\alpha^l\beta^n\ket{n,\pm l}.
\end{equation}
The parameters $\alpha,\beta$ correspond roughly to the semi-major axis and eccentricity 
of the trajectory of the position expectation value respectively, and the label $\pm$ corresponds to counter-clockwise or clockwise rotation \cite{COLAVITA2005183}. For the component of our state corresponding to particles 1 and 2 having constant separation between position expectation values, we find $\beta=0$ and $\alpha$ is the same as considered in previous sections.

\begin{figure}
    \centering
    \includegraphics[scale=0.5]{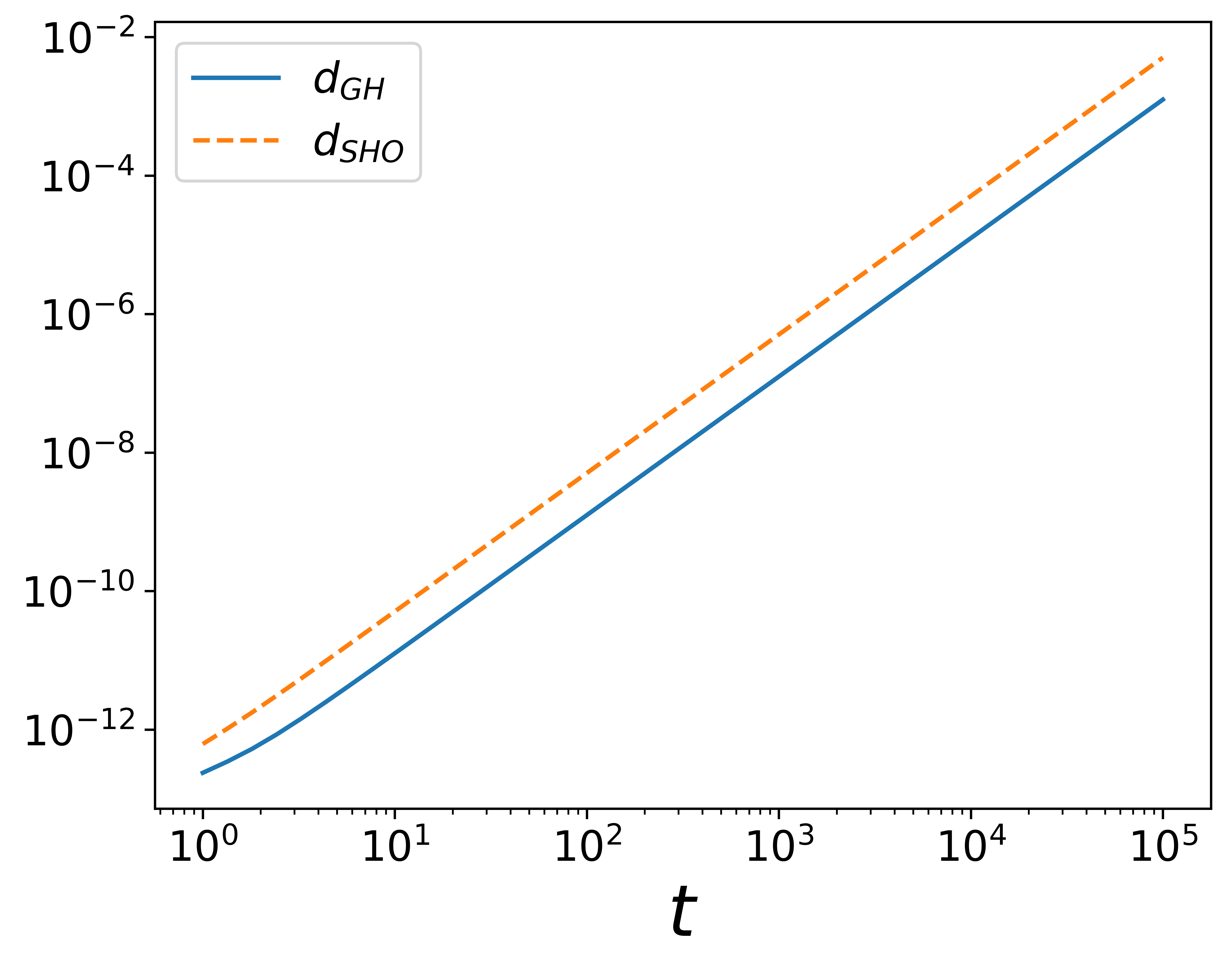}
    \caption{Distance measures $d_{SHO}$ and $d_{GH}$ for various values of the parameters $g_0$ and $\alpha$, with $\frac{g_0}{\alpha}=10^{-6}$ as a function of time (in units $\omega^{-1}$). With $d_{SHO}>d_{GH}$, we have that the numerical approximation is closer to our analytical approximation than to a pure harmonic oscillator as quantified by this distance measure.}
    \label{fig:numerics}
\end{figure}

In order to determine if the numerically approximated time evolved state validates our analytical approximation, we make use of a distance measure between states given by
\begin{equation}
    d(\ket{\psi_1},\ket{\psi_2})=1-\mathrm{Re}\,\bra{\psi_1}\ket{\psi_2}
\end{equation}
and evaluate the distance between our numerical approximation and either the analytical approximation of the  gravitational harmonium, or a pure harmonic oscillator without gravity, both projected onto the subspace spanned by the basis elements we used here, i.e., $d_{GH}\left(U(t)\ket{\alpha},e^{i\delta S}\ket{\alpha(t)}\right)$ and $d_{SHO}\left(U(t)\ket{\alpha},\ket{\alpha(t)}\right)$, respectively. We find that for $1<\alpha<10$ and $\frac{g_0}{\alpha}=10^{-6}$, and $t$ up to $1$s (which corresponds to $t=10^5$ in  units $\omega^{-1}$) as in Sec. \ref{sec:harmonium}, we have $d_{GH}<d_{SHO}$ (see Fig. \ref{fig:numerics}). This establishes that our time evolved state is closer to that of our analytical approximation of the gravitational harmonium than to the evolving harmonic oscillator state with gravity neglected, thus validating our approximation.

\end{document}